\documentclass[aps,pra,showpacs,twocolumn,floats,superscriptaddress]{revtex4-1}
\usepackage{graphicx,dcolumn,longtable,epsfig}
\usepackage[usenames]{color}
\usepackage{amssymb}
\usepackage{amsmath}
\usepackage{epstopdf}
\usepackage{bm}
\usepackage{footnote}
\usepackage{float}
\usepackage{subfigure}
\usepackage{color}
\usepackage[colorlinks=true, citecolor=blue, linkcolor=blue, urlcolor=blue]{hyperref}

\input epsf.tex

\def\bea{\begin{eqnarray}}
\def\eea{\end{eqnarray}}
\def\be{\begin{equation}}
\def\ee{\end{equation}}

\def\prb#1#2#3{Phys.~Rev.~B~{\bf #1},\ #2\ (#3)}

\def\rmp#1#2#3{Rev.~Mod.~Phys.~{\bf #1},\ #2\ (#3)}
\def\prl#1#2#3{Phys.~Rev.~Lett.~{\bf #1},\ #2\ (#3)}
\def\sci#1#2#3{Science~{\bf #1},\ #2\ (#3)}

\def\njp#1#2#3{New~J.~Phys.~{\bf #1},\ #2\ (#3)}

\def\rmp#1#2#3{Rev.~Mod.~Phys.~{\bf #1},\ #2\ (#3)}
\def\jphc#1#2#3{J.~Phys.~C.~{\bf #1},\ #2\ (#3)}

\def\jetp#1#2#3{JETP~{\bf #1},\ #2\ (#3)}

\begin{document}

\author{D. Grimmer}
\affiliation{Homer L. Dodge Department of Physics and Astronomy,
The University of Oklahoma, Norman, Oklahoma ,73019, USA}

\author{A. Safavi-Naini}
\affiliation{Department of Physics, Massachusetts Institute of Technology, Cambridge, Massachusetts, 02139, USA}

\author{B. Capogrosso-Sansone}
\affiliation{Homer L. Dodge Department of Physics and Astronomy,
The University of Oklahoma, Norman, Oklahoma ,73019, USA}

\author{\c{S}.~G. S\"{o}yler}
\affiliation{Max Planck Institute for the Physics of Complex Systems, N\"tohnitzer Stra\ss e 38, 01187 Dresden, Germany}

\title{Quantum Phases of Soft-Core Dipolar Bosons in Optical Lattices}

\begin{abstract}
We study the phase diagram of a system of soft-core dipolar bosons confined to a two-dimensional optical lattice layer. We assume that dipoles are aligned perpendicular to the layer such that the dipolar interactions are purely repulsive and isotropic. We consider the full dipolar interaction and perform Path Integral Quantum Monte Carlo simulations using the Worm Algorithm. Besides a superfluid phase, we find various solid and supersolid phases. We show that, unlike what was found previously for the case of nearest-neighboring interaction, supersolid phases are stabilized not only by doping the solids with particles but with holes as well. We further study the stability of these quantum phases against thermal fluctuations. Finally, we discuss pair formation and the stability of the pair checkerboard phase formed in a bilayer geometry, and suggest experimental conditions under which the pair checkerboard phase can be observed. 
\end{abstract}

\pacs{}

\maketitle



\section{Introdcution}

The recent experimental progress in trapping and controlling polar molecules ~\cite{hetmolecules1,hetmolecules2}, atoms with large magnetic moments~\cite{chromium,erbium,dysprosium} and Rydberg atoms~\cite{rydberg1,rydberg2} has paved the way for the realization of many body quantum systems featuring dipolar interactions. Dipolar interactions, which are long-ranged and anisotropic, have been shown to stabilize a variety of exotic quantum phases such as the supersolid phase. The existence of this phase in solid Helium was debated for a long time partly due to the lack of a solid experimental confirmation~\cite{supersolid1,supersolid2}. Theoretically, it was suggested that supersolidity in solid Helium is due to the presence of a network of dislocations which supports flow~\cite{prokofev}. A recent experimental observation 
of mass transport in solid Helium supports this scenario~\cite{hallock1,hallock2}.  
 
Optical lattice simulators, using ultra-cold atoms and molecules, provide an alternate and promising setup for the observation of the supersolid phase. The tunability and flexibility of these setups allow one to realize systems in which the long-range dipolar interactions are of considerable strength, which can be used to stabilize many novel quantum phases including a supersolid. Accurate and unbiased theoretical predictions, such as those presented in this work, will be crucial in guiding experimentalists in their search for the exotic phases stabilized in these systems. 

In this work, we study a system of soft-core, dipolar bosons confined to a quasi two-dimensional layer, with further confinement provided by a two-dimensional lattice within the layer. We assume that dipoles are aligned perpendicular to the layer by an external field such that the long-range dipolar interactions are purely repulsive and isotropic. Our results are based on Path Integral Quantum Monte Carlo (QMC) using the Worm Algorithm~\cite{Worm}. Previous QMC studies have imposed a cutoff on the dipolar interaction, limiting its range to the nearest neighbors. For example, in Ref.~\cite{Troyer} the authors find that by doping the checkerboard (CB) solid with particles a supersolid phase is stabilized. On the other hand, upon doping with holes, a discontinuous phase transition to a superfluid (SF) was found. 
Similarly, in Ref~\cite{Kawashima}, the authors report a discontinuous phase transition below half filling and away from the tip of the first CB lobe at half filling. 

In the following we consider the full dipolar interaction and show that the discontinuous phase transition from CB to SF is replaced by a continuous transition to a SS phase. Section~\ref{sec:H} describes the system Hamiltonian. Section~\ref{sec:pdt0} presents the ground state phase diagram obtained using the Worm algorithm. Section~\ref{sec:pdtfin} summarizes the finite-temperature behavior of the system. Section~\ref{sec:bilayer} presents our study of pair formation in a bilayer system, where we determine the dipolar interaction strength required to stabilize a pair CB solid at half-filling, as a function of the separation between the layers. Finally, section~\ref{sec:conc} concludes.

\section{System Hamiltonian: Single Layer}\label{sec:H}

In the single band approximation, the system of soft-core dipolar bosons confined to a two-dimensional square lattice is described by the extended Bose Hubbard model
\begin{align}
\label{eq:Hsingle}
\nonumber H&= -t \sum_{\langle i j \rangle} { a_i^\dag a_j^{\,}} + \frac{U}{2} \sum_{i} {n_i(n_i-1)}\\
& +V_{dd}\sum_{i,j}\frac{1}{r_{i,j}^3}n_in_j ,
\end{align}
where $a_i^\dag(a_i^{\,})$ are the boson creation (annihilation) operators following the usual boson commutation relations, $i,\;j$ refer to the lattice sites, $\langle i j \rangle$ denotes nearest-neighboring sites, and $n_i=a_i^\dag a_i$ is the density operator. Here $t$ is the hopping matrix element, $U$ is the on-site, interparticle repulsion.
$V_{\rm dd}=\frac{d^2}{\epsilon_0}$ ($V_{\rm dd}=\mu_0d^2$) is the strength of the electric (magnetic) dipolar interactions where $d$ is the electric (magnetic) dipole moment, and $r_{i,j}=\vert \vec r_i - \vec r_j \vert$ is the separation between two particles on lattice sites $\vec r_i$ and $\vec r_j$. When dipoles are aligned perpendicular to the layer, $V_{\rm dd}$ is purely repulsive and isotropic. 

In the following we present accurate theoretical results based on Path Integral QMC simulations using the Worm Algorithm \cite{Worm}. We have performed the simulations on an $L \times L = N_s$ square lattice system with $L=$8, 12, 16, 20, and 24 and lattice constant $a$. We have imposed periodic boundary conditions in both spatial dimensions. Unless otherwise noted, we use Ewald summation to calculate the full, long-range dipolar interaction.

\section{Ground state phase diagram}\label{sec:pdt0}

\begin{figure}
\begin{center}
\includegraphics[width=0.5\textwidth]{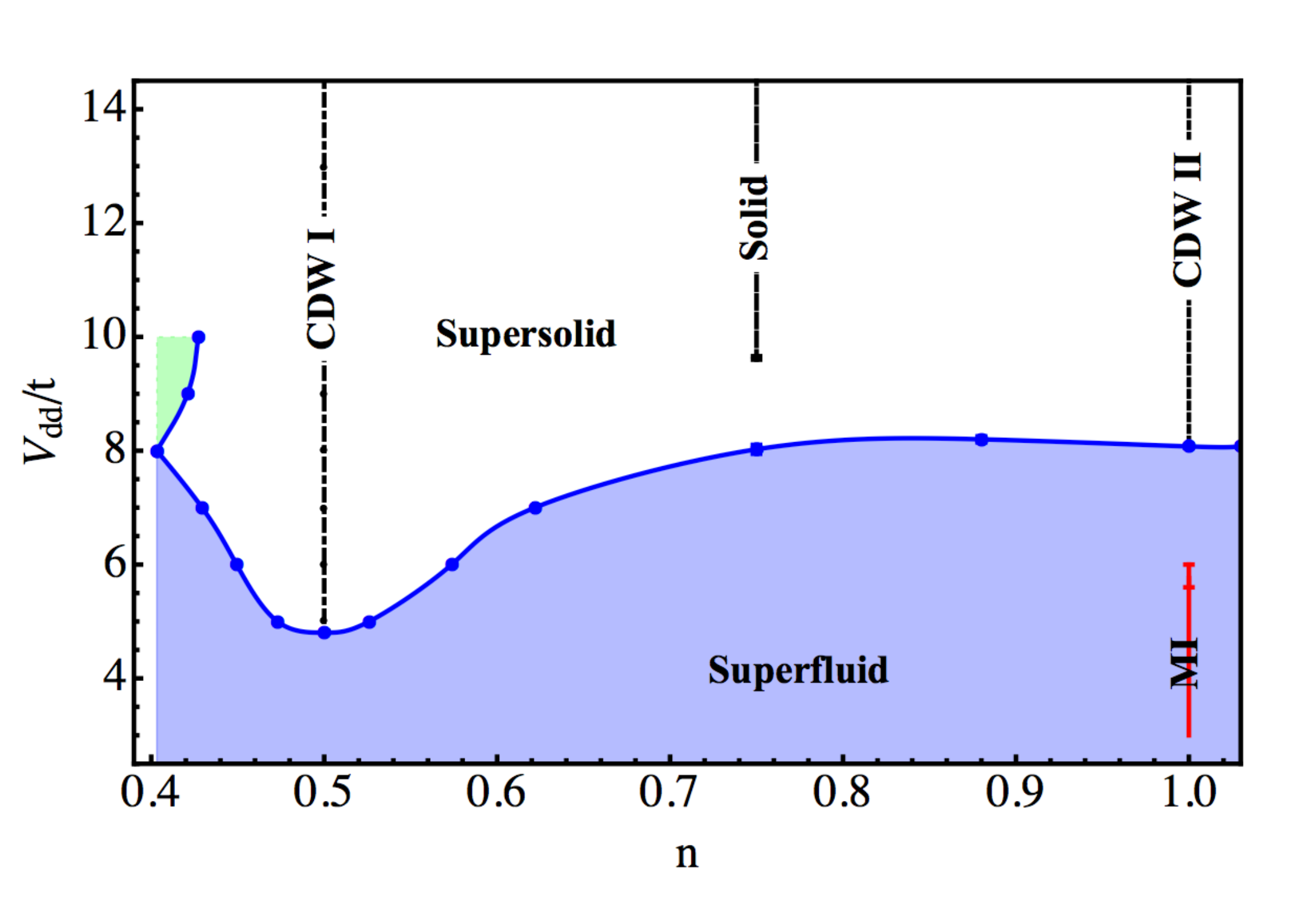}
\caption{(Color online) Phase diagram of Hamiltonian~\eqref{eq:Hsingle} as a function of $V_{\rm dd}/J$ and particle density $n$, computed via QMC simulations, at $U/t=20$ (see text). CDW I, CDW II: Charge density waves at $n=0.5$ and $n=1$, and MI: Mott insulator. The green region corresponds to solids stabilized at different rational fillings.}
\label{fig:PDT0}
\end{center}
\end{figure}

In this section we present the ground state phase diagram of the system described by the Hamiltonian in Eq.~\ref{eq:Hsingle}. Here we have set the onsite interaction strength to $U/t=20$ as done in Ref.~\cite{Troyer}, ensuring the stability of supersolid phase. We have explored the parameter space given by $2<V_{\rm dd}/t<10$ and $0<n\sim1$. 
Fig.~\ref{fig:PDT0} shows the phase diagram of the system as a function of dipolar interaction strength $V_{dd}/t $ and particle density $n=N/N_s$, where $N$ is the number of particles in the system, and $N_s$ is the total number of sites. 

For low enough dipolar interaction $V_{dd}/t \lesssim 5$ and $n\ne1$, the system is in a superfluid state, which is associated with the presence of off-diagonal long-range order. This is characterized by a non-zero, single particle condensate order parameter $\langle \psi\rangle=\langle a_i\rangle \ne 0$ and is associated with a finite value of superfluid stiffness $\rho_{S}=T\langle \mathbf{W}^2 \rangle /d L^{d-2}$ where $\mathbf{W}$ is the winding number in space~\cite{winding}. The superfluid stiffness is directly related to the single particle condensate, and can be calculated within Path Integral Monte-Carlo.

At half filling, upon increasing the dipolar repulsion, the system forms a charge density wave. The charge density wave, which is indicated as CDW I on the phase diagram, is the conventional CB solid, where particles occupy every other lattice site. The CB solid is stabilized due to the repulsive nature of the dipolar interaction. The checkerboard order breaks a discrete $Z_2$ symmetry and is characterized by a finite value of the static structure factor $S(\mathbf{k})$ at the reciprocal lattice vector $\mathbf {k}=(\pi, \pi)$, with
\begin{equation}
S(\mathbf{k})=\frac{1}{N}\sum_{r, r\prime} \exp[i\mathbf{k}(\mathbf{r}-\mathbf{r}^\prime)]\langle n_r n_{r^\prime} \rangle.
\end{equation} 

In the solid phase the system displays zero superfluidity. Since the soft-core inter-particle interaction favors delocalization, the CB solid is stabilized at a higher value of interaction strength compared to the hard-core model, where it was shown that the same phase is stable at $V_{\rm dd}/t \sim 3.6$ \cite{Capogrosso2010}. It should be noted that unlike the results reported in Ref.~\cite{Kawashima} we did not find any evidence of a supersolid at half-filling. Using an interaction cutoff of three nearest neighboring sites, and changing the interaction strength in increments of $\sim 1.3\%$ we were not able to detect this phase. We did not further study the nature of the transition as it was beyond the scope of this work.

Upon doping the system with particles or holes, i.e. moving along the horizontal direction from the CB phase, we enter the SS phase which displays both broken translational symmetry, i.e. $S(k)\neq0$, and off-diagonal long-range order, i.e. $\rho_S\ne 0$. For $V_{dd}/t\lesssim8$, further increasing (decreasing) the particle density above (below) some critical filling, destroys the translational order via a second order phase transition belonging to the $(2+1)$ Ising universality class, leaving the system in an SF phase. The SF phase can also be reached at fixed density by decreasing the dipolar interaction strength. The boundary between the SS and SF phases is found by using standard finite size scaling. Specifically, we determine critical points using finite size scaling for the static structure factor by plotting $S(\pi,\pi) L^{2\beta/\nu}$ vs. $n$ or $V_{dd}/t$, with scaling coefficient $2\beta/\nu=1.0366$~\cite{Ising}. Critical points are determined from the intersection of $S(\pi,\pi) L^{2\beta/\nu}$ curves for different $L$'s. 
The boundary forms a lobe-like structure which is asymmetric due to the lack of  particle-hole symmetry. 

\begin{figure}
\begin{center}
\includegraphics[width=0.5\textwidth]{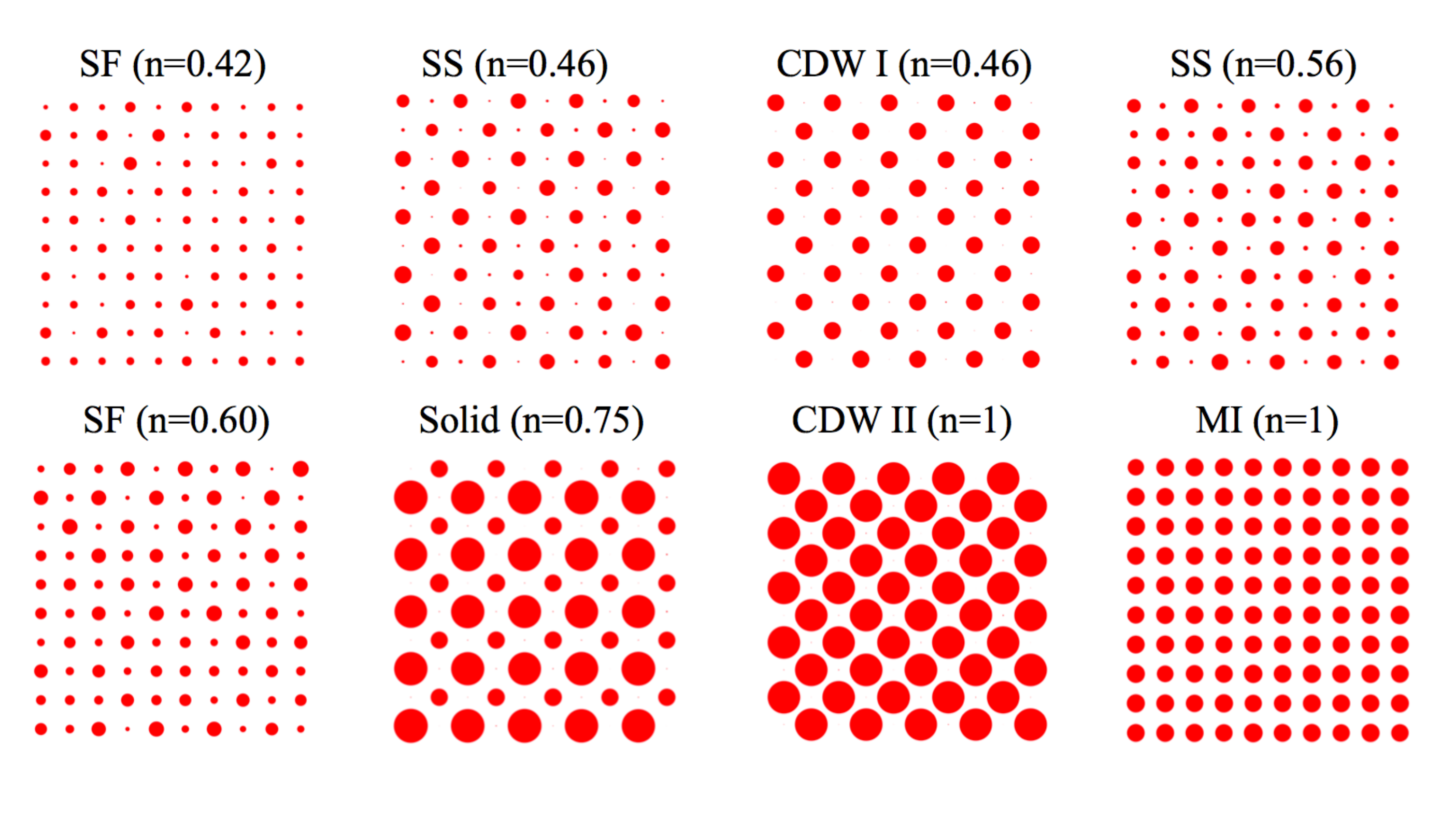}
\caption{(Color online) Configurations corresponding to the phases stabilized by model Eq.~\ref{eq:Hsingle} as shown in Fig.~\ref{fig:PDT0}. The radius of each sphere is proportional to the density at a given site for a specific Monte Carlo configuration.}
\label{fig:phaseschem}
\end{center}
\end{figure}

For higher dipolar interaction strengths, further doping with holes results in the formation of various solid states corresponding to different rational fillings (see also the hard core case~\cite{Capogrosso2010}). This is indicated by the green shaded area in Fig.~\ref{fig:PDT0}. 
On the other hand, on the particle doping side, SS phase extends all the way to filling factor $n=1$ with one exception. At $n=0.75$ a new solid phase is stabilized. This phase is composed of two square sub-lattices with doubled unit cell and filling factors 1 and 2 (see Fig.~\ref{fig:phaseschem}), hence breaking a further $Z_2$ symmetry. Finally a third solid (CDW II) is formed at $n=1$. This is another CB phase, where we have double occupancy in CB order. While all the CB solids are surrounded by SS, the structure of the SS phase differs depending on whether it is in the vicinity of the solid at $n=1/2$, $n=3/4$, or $n=1$. For instance, on the left of CDW II the SS phase is the result of coherent hole excitations over CDWII solid. Configurations corresponding to the phases stabilized by model Eq.~\ref{eq:Hsingle} are shown in Fig.~\ref{fig:phaseschem}. The radius of each sphere is proportional to the density at a given site for a specific Monte Carlo configuration.

At integer filling factor $n=1$ and for low enough dipolar interaction the system is in a Mott Insulator (MI) phase (indicated by the red line in Fig.~\ref{fig:PDT0}). Upon increasing the dipolar interaction at fixed unit filling, particle delocalization is favored and the system undergoes a second order phase transition in favor of a SF phase. This is the standard U(1) MI-SF transition in (2+1) dimensions. Clearly, the gapped MI can also be destroyed by doping away from integer filling as in the standard generic MI-SF transition.
Further increase of the dipolar interaction at unit filling results in the formation of a checkerboard solid (CDWII) at $V_{\rm dd}/t=8.075\pm 0.025$. We were not able to resolve the nature of transition within our statistical error. It is worth noting that the behavior of the system at unit filling differs from what found in Ref.~\cite{Troyer} where a direct MI-CDWII discontinuous transition was found at $n=1$.
  \begin{figure}
\begin{center}
\includegraphics[width=0.4\textwidth]{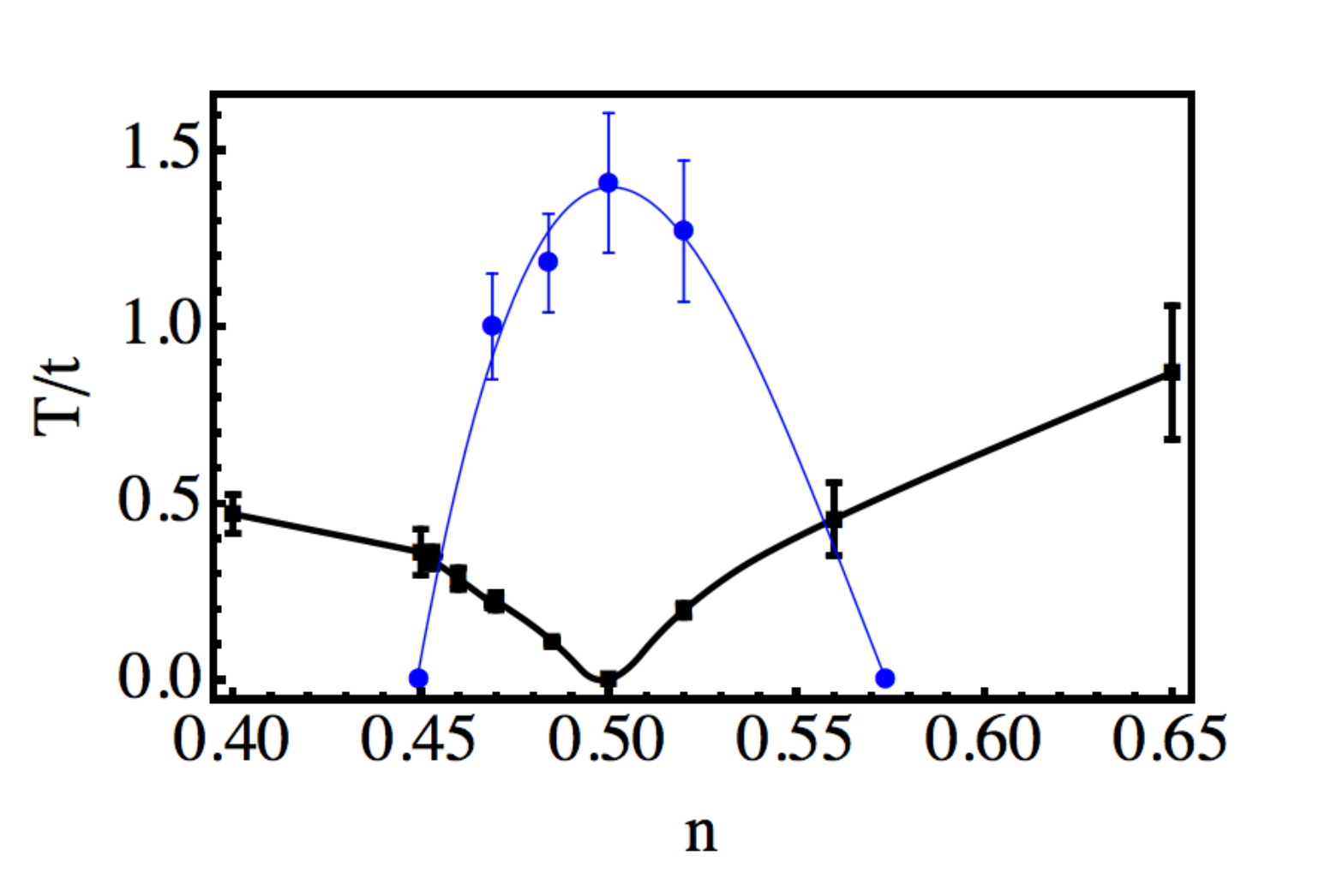}
\caption{(Color online) Critical temperatures for the disappearence of off-diagonal long-range order (black) and diagonal long-range order (blue) for $V_{\rm dd}/t=6$. The SS phase disappears via a two-step transition (see text for details). }
\label{fig:PDT}
\end{center}
\end{figure}

\section{Finite temperature results}\label{sec:pdtfin}
We have studied the stability of the quantum phases described above against thermal fluctuations. As an example, we show our results for $V_{\rm dd}/t=6$ which are summarized in Fig.~\ref{fig:PDT}. While superfluidity disappears via a Kosterlitz-Thouless (KT) type~\cite{KTtrans} transition, the CB solid melts via a two-dimensional Ising-type transition with $2\beta/\nu=1/4$. In the case of the SS phase, the disappearance happens via a two-step process. In other words, there exist two critical temperatures $T_{\rm KT, SS}$ and $T_{\rm CB, SS}$ corresponding to the disappearance of off-diagonal and diagonal long-range order respectively. Depending on the density, $T_{\rm KT, SS}\gtrless T_{\rm CB, SS}$. As shown in Fig.~\ref{fig:PDT}, on the approach of the SS-SF transition at zero temperature, the diagonal order disappears at lower temperatures than the off-diagonal one. On the other hand, at densities close enough to half filling the off-diagonal order disappears first and the SS phase melts into a liquid-like phase reminiscent of a liquid crystal.

\begin{figure}
\begin{center}
\includegraphics[width=0.4\textwidth]{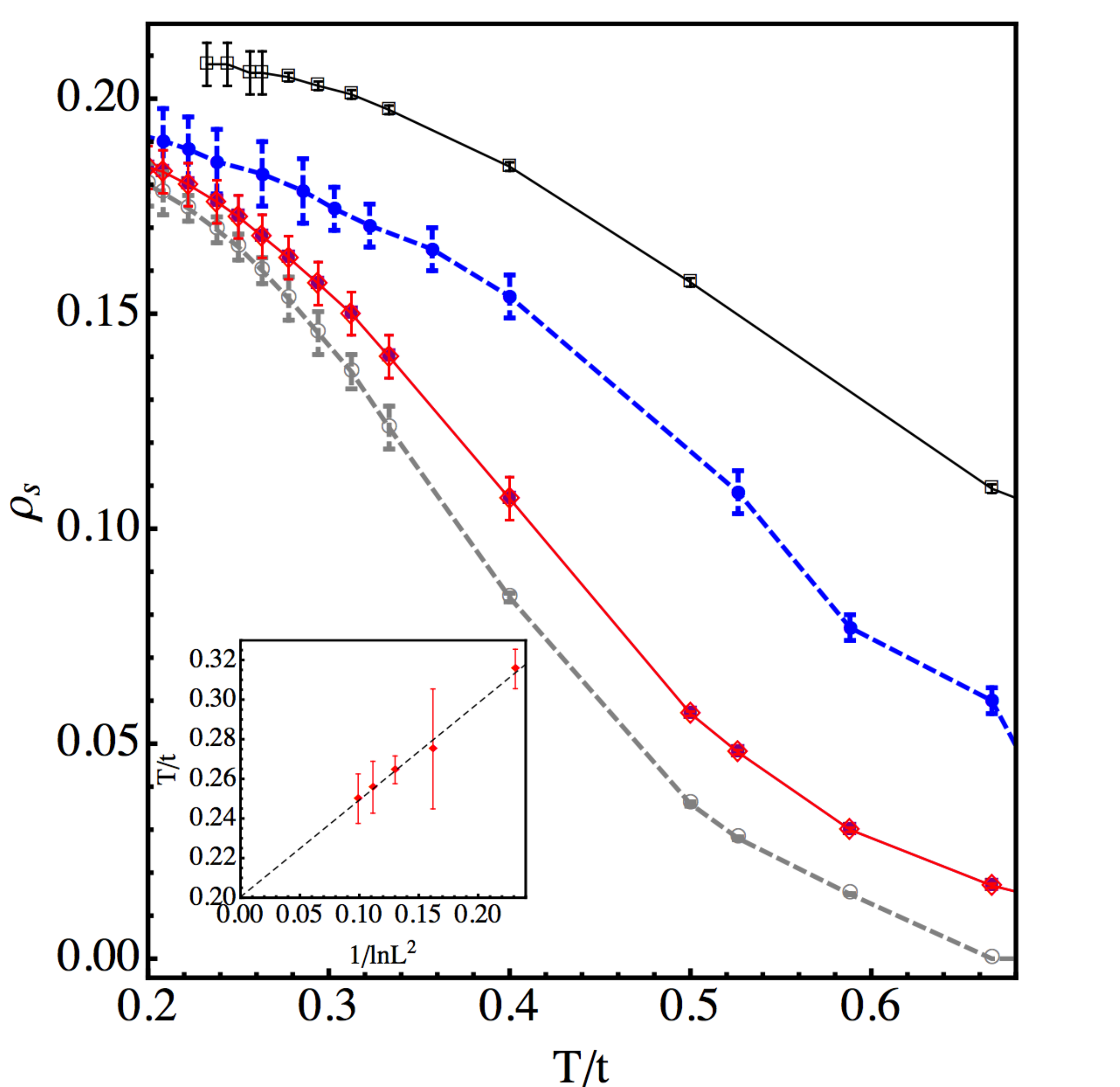}
\caption{(Color online) The superfluid stiffness $\rho_s$ vs. temperature $T/t$ at $V/t=6$ and $n=0.469$ for $L=$ 8, 12, 16, 20, and 24 shown using open squares, filled circles, filled squares, open circles, and filled stars respectively. The supersolid phase melts in two stages. The inset shows the Kosterlitz-Thouless scaling described in~\cite{Ceperley2} used to find the critical temperature for the first transition. At $T_{\rm KT, SS}\approx 0.20\pm 0.01$ the SS melts into a liquid-crystal like phase. The critical temperature is given by the intersection of dashed line of best fit with the vertical axis.  }
\label{fig:finT}
\end{center}
\end{figure}

Fig.~\ref{fig:finT} illustrates an example of finite size scaling used to determine $T_{\rm KT}$. The main panel shows the superfluid density $\rho_s$ vs. $T/t$ for different system sizes at $V_{\rm dd}/t=6$ and $n=0.469$. The inset shows the finite size scaling procedure described in ~\cite{Ceperley2} where the dashed line is a linear fit, and the intersection with the vertical axis determines the critical point, $T_{\rm KT,SS}\approx 0.20\pm 0.1$. 


\section {Checkerboard solid in a bilayer geometry}\label{sec:bilayer}
In analogy with our previous work reported in~\cite{bilayers}, we have studied pair formation and the stability of the pair checkerboard (PCB) phase in a bilayer system. In this geometry, the interlayer dipolar interaction possesses an attractive component. The interlayer interactions is purely attractive if the two dipoles are directly on top of one another. In this case the interlayer interaction takes the form $V_{\rm dd}^{\perp}/t=-2V_{\rm dd}/d_z^3$ and favors the pairing of dipoles sitting on top of each other. The ratio between attractive and repulsive interaction can be tuned by changing the distance $d_z$ between layers. The PCB phase is similar to the conventional CB phase, in that the atoms in each layer occupy every other site of the lattice. As a result, the PCB phase is characterized by a finite value of the static structure factor $S(\pi, \pi)$. Additionally, in this phase, the atoms across the layers are strongly paired, which results in strong correlations in the positions of the two checkerboard solids. 
\begin{figure}
\begin{center}
\includegraphics[width=0.5\textwidth]{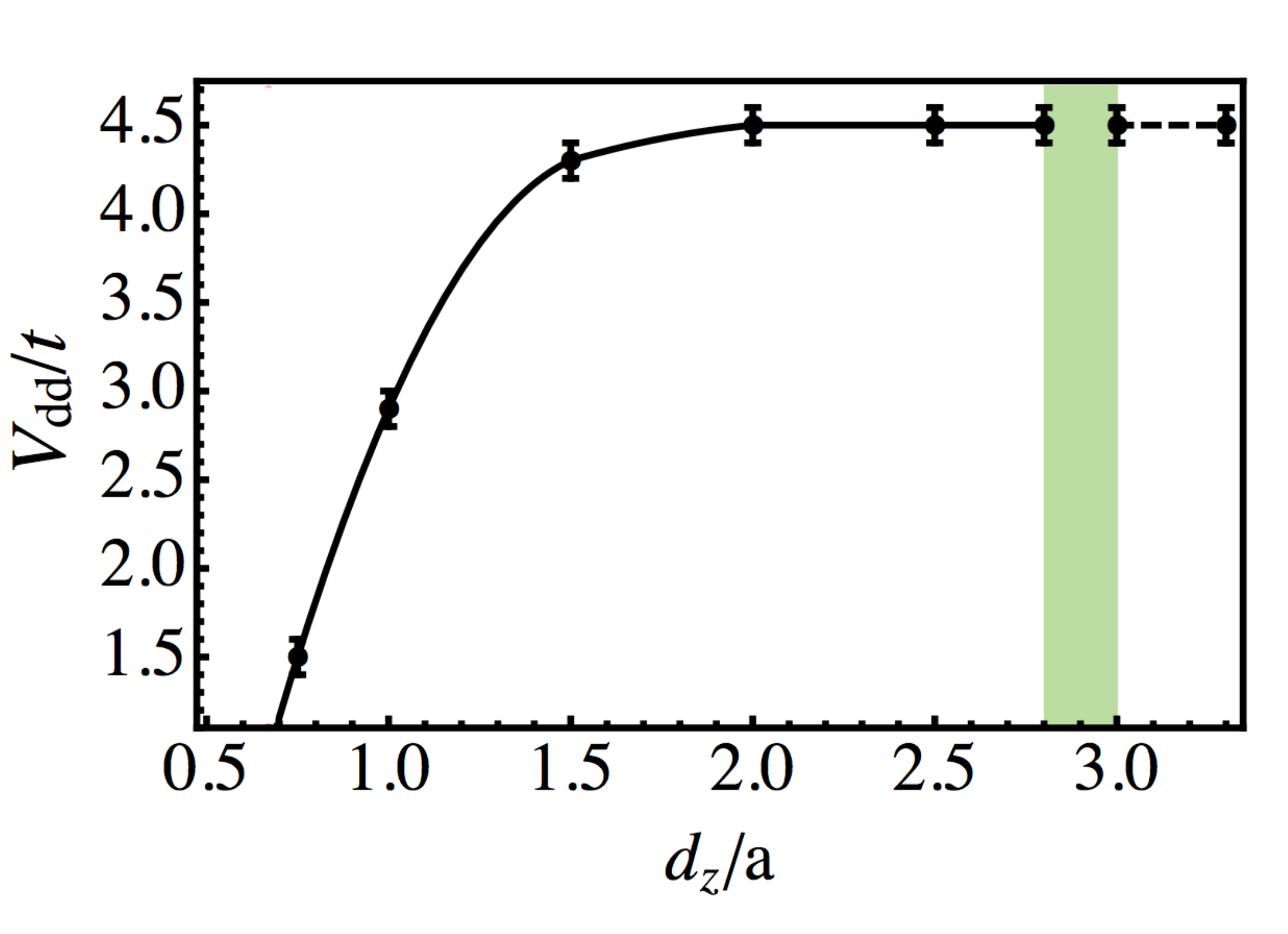}
\caption{(Color online) 
Plot of the minimum value $V_{\rm dd}/t$ needed to stabilize the CB phase in the case of a bilayer geometry as a function of the distance between the layers $d_z/a$. Once the layers are separated by $d_z/a > 2.8$ they behave as independent layers, that is the CB solids on the two layers are not correlated and the minimum $V_{\rm dd}/t$ has saturated to the case of a single layer. Note that these results where obtained using a cutoff of three nearest neighbors. This results in a shift of $\sim 7\%$ of the minimum $V_{\rm dd}/t$ for the single layer.}
\label{fig:CBvsd}
\end{center}
\end{figure}

Fig.~\ref{fig:CBvsd} presents the minimum dipolar interaction strength $V_{\rm dd}/t$ required to form a CB solid in a bilayer system as a function of different values of interlayer separation $d_z/a$. For computational convenience, we have used a cut-off of three nearest neighboring sites for dipolar interaction range. This results in a shift of $\sim 7\%$ of the minimum $V_{\rm dd}/t$ for the single layer. In order to establish whether the solid phase is paired we have performed several simulations with different initial conditions for each set of parameters and observed whether the equilibrium configuration was dependent on the initial choice or not. Fig.~\ref{fig:CBvsd} shows that in the regime where $V_{\rm dd}^{\perp}/t \gg V_{\rm dd}/t$, corresponding to small inter-layer separations, the PCB phase is stabilized at a considerably smaller $V_{\rm dd}/t$ compared to the single layer case presented earlier. This is due to the larger effective mass of the pairs which stabilizes the CB solid at smaller interaction strength. As the separation between the two layers is increased the pairs are destabilized. We indicate the separation beyond which the two layers are uncorrelated using the shaded region.  

The bilayer setup can play an instrumental role in the observation of the quantum phases stabilized by dipolar interaction. The interlayer attraction which leads to pairing, creates a higher effective mass for the particle forming the paired phases. This in turn allows one to form phases like PCB at lower interaction strengths compared to the single layer CB. Correspondingly, at a given interaction strength where both CB and PCB have been stabilized in the ground state, the PCB phase is more robust against thermal fluctuations, resulting in higher melting temperatures. 
Here we present experimental estimates for the conditions under which the PCB phase can be observed. For example, with a gas of Dy ($d=10 \mu_B$) a choice of lattice parameters $a=250$~nm, $d_z=200$~nm, $J=\,50 h$Hz stabilizes the PCB phase with $V_{dd}/J\sim 1 $. Similarly using a gas Er$_2$ Feshbach molecules~\cite{Ferlaino, Dalmonte2010} ($d=14 \mu_B$) with $a=300$~nm, $d_z=200$~nm, $J=100\, h$Hz the PCB phase is stabilized at $V_{dd}/J\sim 0.4$. In both cases the PCB phase can be observed at nk temperatures.

Using RbCs ($d=0.3$D) and typical trapping parameters $a=500$~nm, $d_z=300$~nm and $J=150\, h$Hz we find $V_{dd}/J\sim 0.75$, which is large enough to stabilize the PCB. The latter survives up to $T_c^{PCB}\sim 10$~nK.

\section{Conclusions}\label{sec:conc}
We have presented results for the phase diagram of a system of soft-core bosons confined to a two-dimensional optical lattice layer.  Particles are interacting via an isotropic dipolar repulsive interaction. We have performed Path Integral Quantum Monte Carlo simulations using the Worm Algorithm at fixed strength of the onsite interaction. Besides a superfluid phase, we have found various solid and supersolid phases. In particular we have found checkerboard density waves at fillings $n=0.5,0.75,1$. We have shown that, unlike what was found previously for the case of nearest-neighboring interaction, supersolid phases are stabilized not only by doping the solids with particles but with holes as well. Moreover, we find that at unit density a superfluid phase intervenes in between the Mott insulator, stabilized at lower dipolar interaction, and the charge density wave consisting of alternating empty and doubly occupied sites, stabilized at larger dipolar interaction. This too is in contrast with previous findings. We have further studied the stability of these quantum phases against thermal fluctuations. Finally, we have discussed pair formation and the stability of the pair checkerboard phase formed in a bilayer geometry, and suggested experimental conditions under which the pair checkerboard phase can be observed.

\emph{Acknowledgments} This work used the Extreme Science and Engineering Discovery Environment (XSEDE), which is supported by National Science Foundation grant number ACI-1053575.

\vspace*{-5mm}


\begin{thebibliography}{50}
\vspace*{-5mm}
\bibitem{hetmolecules1} K.-K. Ni, S. Ospelkaus, M. G. H. de Miranda, A. PeÕer, B. Neyenhuis, J. J. Zirbel, S. Kotochigova, P. S. Julienne, D. S. Jin, and J. Ye, \sci{322}{231}{2008}. 
\bibitem{hetmolecules2} C.-H. Wu, J. W. Park, P. Ahmadi, S. Will, and M. W. Zwierlein, \prl{109}{085301}{2012}.
\bibitem{chromium}A. Griesmaier, J. Werner, S. Hensler, J. Stuhler, and T. Pfau, \prl{94}{160401}{2005}.
\bibitem{erbium} K. Aikawa, A. Frisch, M. Mark, S. Baier, A. Rietzler, R. Grimm, and F. Ferlaino, \prl{108}{210401}{2012}.
\bibitem{dysprosium} M. Lu, N. Q. Burdick, S. H. Youn, and B. L. Lev, \prl{107}{190401}{2011}.
\bibitem{rydberg1} M. Saffman, T. G. Walker and K. M\o lmer,\rmp{82}{2313}{2010}.
\bibitem{rydberg2}  D. Comparat and Pillet, J. Opt. Soc. Am. B \textbf{27}, A208-A232 (2010).
\bibitem{supersolid1} A. J. Leggett, \prl{25}{1543}{1970}.
\bibitem{supersolid2} M. Boninsegni and N. Prokof'ev \rmp{84}{759}{2012}.
\bibitem{prokofev} N. Prokof'ev, Adv. Phys. \textbf{56}, 381 (2007).
\bibitem{hallock1}M. W. Ray and R. B. Hallock, \prl{105}{145301}{2010}.
\bibitem{hallock2} Ye Vekhov, W.J. Mullin, and R.B. Hallock, \prl{113}{035302}{2014}.
\bibitem{Worm}N. V. Prokof'ev, B. V. Svistunov and I. S. Tupitsyn \prl{238}{ 253}{1998}; \jetp {87}{310}{1998}.
\bibitem{Troyer} P. Sengupta, L. P. Pryadko, F. Alet, M. Troyer, and G. Schmid, \prl{94}{207202}{2005}.
\bibitem{Kawashima} T.  Ohgoe, T. Suzuki, N. Kawashima, \prb{86},{054520}{2012}.
\bibitem{winding}E. L. Pollock and D. M. Ceperley \prb{36}{8343}{1987}.
\bibitem{Capogrosso2010} B. Capogrosso-Sansone, C. Trefzger, M. Lewenstein, P. Zoller and G. Pupillo \prl{104}{125301}{2010}.
\bibitem{Ising}M. Hasenbusch, K. Pinn and S. Vinti \prb{59}{11471}{1999}.
\bibitem{KTtrans} J. M. Kosterlitz and D. J. Thouless \jphc {6}{1973}{1181}.
\bibitem{Ceperley2}D. M. Ceperley and E. L. Pollock \prb{39}{2084}{1989}.
\bibitem{bilayers} A. Safavi-Naini, S. G. Soyler, G. Pupillo, H. R. Sadeghpour, B. Capogrosso-Sansone, \njp{15}{013036}{2013}. 
\bibitem{Ferlaino}K. Aikawa, A. Frisch, M. Mark, S. Baier, A. Rietzler, R. Grimm, and F. Ferlaino, \prl{108}{210401}{2012}.
\bibitem{Dalmonte2010}M. Dalmonte, G. Pupillo and P. Zoller \prl{105}{140401}{2010}.
\end{thebibliography}
\end{document}